\documentclass[11pt]{article}
\usepackage{amsmath,amsthm}

\newcommand{\R}{{\mathbf R}} \newcommand{\N}{{\mathbf N}}
\newcommand{\K}{{\mathbf K}} \newcommand{\Z}{{\mathbf Z}}
  \def\C{{\mathbf C}}
\newcommand{\Prm}{{\mathbf P}}
\newcommand{\wt}{\widetilde }

\renewcommand{\epsilon}{\varepsilon } 
\newcommand{\g}{\gamma } 
\renewcommand{\rho}{\varrho } 
\renewcommand{\l}{{\lambda }} 
\renewcommand{\phi}{\varphi }
\renewcommand{\a}{\alpha }
\renewcommand{\b}{\beta }

\def\rs{\right>}
\def\lg{\left|}

\newtheorem{theorem}{Theorem}
\newtheorem{lemma}{Lemma}
\newtheorem{corollary}{Corollary}
\newtheorem{proposition}{Proposition}

\begin {document}
 \title{Quantum Integration in Sobolev Classes}

\author {Stefan Heinrich\\
Fachbereich Informatik\\
Universit\"at Kaiserslautern\\
D-67653 Kaiserslautern, Germany\\
e-mail: heinrich@informatik.uni-kl.de\\
homepage: http://www.uni-kl.de/AG-Heinrich}   
\date{}
\maketitle

\date{}
\maketitle
\begin{abstract}
We study high dimensional integration in the quantum model of 
computation. We develop quantum algorithms for integration of functions from Sobolev classes
$W^r_p([0,1]^d)$ and  analyze their convergence rates. We also prove lower bounds which show
that the proposed algorithms are, in many cases, optimal within the setting of quantum computing.
This extends recent results of Novak on integration of functions from H\"older classes. 
\end{abstract}
\section{Introduction}
Since Shor's (1994) discovery of a polynomial factoring algorithm on a 
quantum computer, the question of the potential 
power of quantum computing was posed and studied for many problems 
of computer science. Most of these are of discrete type, while so far 
little was done for numerical
problems of analysis. This field contains a variety of  intrinsically difficult
problems. One of them is high dimensional integration.

To judge possible gains by a quantum computer, one first of all needs to know
the complexity of the respective problem in the classical settings.
The complexity of many basic numerical problems in the classical deterministic
and randomized setting is well understood
due to previous efforts in information-based complexity theory (see 
Traub, Wasilkowski, and Wo\'zniakowski, 1988, Novak, 1988, and Heinrich, 1993). 
This theory
established precise complexity rates by developing optimal algorithms, 
on one hand, and proving matching lower bounds, on the other.

Based on such grounds, it is a challenging task to study these problems in the quantum 
model of computation and compare the results to the known classical complexities, this
way locating problems where quantum computing could bring essential speedups, 
and moreover,
quantitatively assessing the reachable gain. 

In a series of papers, Novak and the author started to investigate this field. 
Their research dealt with summation of sequences and integration of functions.
So Novak (2001) studies integration of functions from H\"older spaces, using the
algorithm of Brassard, H{\o}yer, Mosca, and Tapp (2000) for approximating the mean
of uniformly bounded sequences.  Heinrich (2001a) and Heinrich and Novak (2001b) 
developed quantum algorithms for the mean of $p$-summable sequences and proved
their optimality. Moreover, such an approach required a formal model of quantum 
computation for numerical problems, which was developed and studied in 
Heinrich (2001a). This way the basic elements of a quantum setting of information-based 
complexity
theory were established. First ideas about path integration are discussed in 
Traub and Wo\'zniakowski (2001).

Integration of functions from Sobolev spaces is one of the basic numerical 
problems for which we know the complexity both in the classical deterministic
and randomized setting. In the present paper we study this question in the quantum setting.
We develop a quantum integration
algorithm by splitting the problem into levels, using a hierarchy of quadrature
formulas, and this way reducing it to computing the mean of families of $p$-summable sequences.
This enables us to apply the results of Heinrich (2001a) and Heinrich and Novak (2001b), 
and shows that
the investigation of $p$-summable sequences was an important prerequisite to handle
functions from Sobolev classes.
We also prove lower bound which show the optimality (up to logarithmic factors)
of the proposed algorithms.

The contents of the paper is as follows. In section 2 we recall some notation 
from the quantum setting for numerical problems as developed in Heinrich (2001a).
In section 3 we add some new results of general type which will be needed later on.
Section 4 recalls known facts about summation of sequences and provides some
refinements of estimates. The main result about quantum integration of functions
from Sobolev classes is stated and proved in section 5. 
The paper concludes with section 6 containing comments on the quantum bit model and a
summary including comparisons to the classical deterministic and randomized setting.

For more details on the quantum setting for numerical problems we refer to
Heinrich (2001a), also to the survey by Heinrich and Novak (2001a), and
to an introduction by Heinrich (2001b).   
Furthermore, for general background on quantum computing we refer to the surveys 
Aharonov (1998), Ekert, Hayden, and Inamori (2000), Shor (2000), and to the 
monographs Pittenger (1999), 
Gruska (1999), and Nielsen and Chuang (2000).

\section{Notation}

For nonempty sets $D$ and $K$, we denote by $\mathcal{F}(D,K)$
the set of all functions from $D$ to $K$. Let 
$F \subseteq \mathcal{F}(D, K)$ be a nonempty subset. 
Let $\K$ stand for either 
$\R$ or $\C$, the field of real or complex numbers, let  
$G$ be a normed space over $\K$, and 
let $S:F\to G$ be a mapping. We seek to 
approximate $S(f)$ for $f\in F$ by means of quantum computations. 
Let $H_1$ be the 
two-dimensional complex Hilbert space $\C^2$, with its unit vector
basis $\{e_0,e_1\}$, let
$$
 H_m=\underbrace{H_1\otimes\dots\otimes H_1}_{m}, 
$$
equippeded with the tensor
Hilbert space structure. 
Denote
$$\Z[0,N) := \{0,\dots,N-1\}$$
for $N\in\N$, where we agree to write, as usual, $\N= \{1,2,\dots \}$ 
and $\N_0=\N\cup\{0\}$.
Let $\mathcal{C}_m = \{\lg i\rs:\, i\in\Z[0,2^m)\}$ be the canonical basis of
$H_m$, where  $\lg i \rs$ stands for 
$e_{j_0}\otimes\dots\otimes e_{j_{m-1}}$, $i=\sum_{k=0}^{m-1}j_k2^{m-1-k}$ the binary 
expansion of $i$. Let $\mathcal{U}(H_m)$ denote the set of unitary operators on $H_m$. 

A quantum query  on $F$ is given by a tuple
\begin{equation*}
Q=(m,m',m'',Z,\tau,\beta),
\end{equation*}
where $m,m',m''\in \N, m'+m''\le m, Z\subseteq \Z[0,2^{m'})$ is a nonempty 
subset, and
$$\tau:Z\to D$$
$$\beta:K\to\Z[0,2^{m''})$$
are arbitrary mappings. We let $m(Q):=m$ be the number of qubits of $Q$. 

Given a query $Q$, we define for each $f\in F$ the unitary operator 
$Q_f$ by setting for  
$\lg i\rs\lg x\rs\lg y\rs\in \mathcal{C}_m
=\mathcal{C}_{m'}\otimes\mathcal{C}_{m''}\otimes\mathcal{C}_{m-m'-m''}$:
\begin{equation*}
Q_f\lg i\rs\lg x\rs\lg y\rs=
\left\{\begin{array}{ll}
\lg i\rs\lg x\oplus\beta(f(\tau(i)))\rs\lg y\rs &\quad \mbox {if} \quad i\in Z\\
\lg i\rs\lg x\rs\lg y\rs & \quad\mbox{otherwise,} 
 \end{array}
\right. 
\end{equation*}
where $\oplus$ means addition modulo $2^{m''}$. 

A quantum algorithm  on $F$  with no measurement is a tuple
\begin{equation*}
A=(Q,(U_j)_{j=0}^n).
\end{equation*}
Here $Q$ is a quantum query on $F$, $n\in\N_0$  and
$U_{j}\in \mathcal{U}(H_m)\,(j=0,\dots,n)$, with $m=m(Q)$.
Given $f\in F$,
we define $A_f\in \mathcal{U}(H_m)$ as
\begin{equation*}
A_f = U_n Q_f U_{n-1}\dots U_1 Q_f U_0.
\end{equation*}
We denote by $n_q(A):=n$ the number of queries and by $m(A)=m=m(Q)$ the 
number of qubits of $A$. Let $(A_f(x,y))_{x,y\in \Z[0,2^m)}$ 
be the matrix of the 
transformation $A_f$ in the canonical basis $\mathcal{C}_{m}$.

A quantum algorithm from $F$ to $G$ with $k$ measurements  is a tuple
$$
A=((A_\ell)_{\ell=0}^{k-1},(b_\ell)_{\ell=0}^{k-1},\varphi),
$$ 
where $k\in\N$, $A_\ell\;(\ell=0,\dots,k-1)$ are quantum algorithms
on $F$ with no measurements, 
$$
b_0\in\Z[0,2^{m_0}), 
$$
$$
b_\ell:\prod_{i=0}^{\ell-1}\Z[0,2^{m_i}) \to \Z[0,2^{m_\ell})\quad 
(1\le \ell \le k-1),
$$
where $m_\ell:=m(A_\ell)$, and 
$$
\varphi:\prod_{\ell=0}^{k-1}\Z[0,2^{m_\ell}) \to G.
$$
The output of $A$ at input $f\in F$ will be a probability measure $A(f)$ on $G$, 
defined as follows: First put
\begin{eqnarray}   
p_{A,f}(x_0,\dots, x_{k-1})&=&
|A_{0,f}(x_0,b_0)|^2 |A_{1,f}(x_1,b_1(x_0))|^2\dots\nonumber\\
&&\dots |A_{k-1,f}(x_{k-1},b_{k-1}(x_0,\dots,x_{k-2}))|^2.\nonumber
\end{eqnarray}
Then define $A(f)$ by setting for any subset $C\subseteq G$
\begin{equation*}
A(f)(C)=\sum_{\phi(x_0,\dots,x_{k-1})\in C}p_{A,f}(x_0,\dots, x_{k-1}).
\end{equation*}
Let 
$n_q(A):=\sum_{\ell=0}^{k-1} n_q(A_\ell)$
denote the number of queries used by $A$.
For more details and background see Heinrich (2001a).
Note that we often use the term `quantum algorithm' (or just `algorithm'), 
meaning a quantum algorithm with measurement(s).

If $A$ is an algorithm with one measurement, the above definition 
simplifies essentially. Such an algorithm is given by
$$
A=(A_0,b_0,\phi),\quad A_0=(Q,(U_j)_{j=0}^n).
$$
The quantum computation is carried out on $m:=m(Q)$ qubits. For $f\in F$ the 
algorithm starts in the state  $\lg b_0 \rs$ and produces
$$
\lg \psi_f \rs=U_n Q_f U_{n-1}\dots U_1 Q_f U_0\lg b_0 \rs.
$$
Let
$$ 
\lg \psi_f \rs=\sum_{i=0}^{2^m-1}a_{i,f}\lg i \rs
$$
(referring to the notation above, we have $a_{i,f}=A_{0,f}(i,b_0)$). Then $A$ outputs the 
element $\phi(i)\in G$ with probability $|a_{i,f}|^2$. It is shown in Heinrich (2001a),
Lemma 1, that for each algorithm $A$ with $k$ measurements there is an algorithm 
$\wt{A}$ with one measurement such that $A(f)=\wt{A}(f)$ for all $f\in F$ and 
$\wt{A}$ uses just twice the number of queries of $A$, that is, $n_q(\wt{A})=2n_q(A)$.
Hence, as long as we are concerned with studying minimal query error and complexity
(see below) up to the 
order, that is, up to constant factors, we can 
restrict ourselves to algorithms with one measurement.

Let $\theta\ge 0$. For a quantum algorithm $A$ we 
define the (probabilistic) error at $f\in F$ as follows. 
Let 
$\zeta$ be a random variable with distribution $A(f)$. Then 
\begin{equation*}
e(S,A,f,\theta)=\inf\left\{\varepsilon\ge0\,\,|\,\,\Prm\{\|S(f)-\zeta\|>\varepsilon\}\le\theta
\right\}
\end{equation*}
(note that this infimum is always attained).
Hence $e(S,A,f,\theta)\le \varepsilon$ iff the algorithm $A$ computes $S(f)$  with 
error at most $\varepsilon$ and probability at least $1-\theta$. 
Trivially, $e(S,A,f,\theta)=0$ for $\theta\ge 1$. 
We put 
$$
e(S,A,F,\theta)=\sup_{f\in F} e(S,A,f,\theta) 
$$
(we allow the value $+\infty$ for this quantity). Furthermore, we set
\begin{eqnarray*}
\lefteqn{e_n^q(S,F,\theta)}\\
&=&\inf\{e(S,A,F,\theta)\,\,|\,\,A\,\,
\mbox{is any quantum algorithm with}\,\, n_q(A)\le n\}.
\end{eqnarray*}
It is customary to consider these quantities at a fixed error probability
level: We denote
$$
e(S,A,f)=e(S,A,f,1/4)
$$
and similarly,
$$
e(S,A,F)=e(S,A,F,1/4),\quad e_n^q(S,F)=e_n^q(S,F,1/4).
$$
The choice $\theta=1/4$ is arbitrary - any fixed $\theta<1/2$ would do.
The quantity $e_n^q(S,F)$ is  central for our study - it is the 
$n$-th minimal query error, 
that is, the smallest error which can be reached using at most $n$ queries.
Note that it essentially suffices to study
$e_n^q(S,F)$ instead of $e_n^q(S,F,\theta)$, 
since with $\mathcal{O}(\nu)$ repetitions, the error probability 
can be reduced to $2^{-\nu}$ (see Lemma \ref{lem:3} below).
 
The query complexity is defined for $\varepsilon > 0$ by 
\begin{eqnarray*}
\lefteqn{\mbox{comp}_\varepsilon^q(S,F)=}\\
&&\min\{n_q(A)\,\,|\,\, A\,\,\mbox{is any quantum 
algorithm with}\,\, e(S,A,F) \le \varepsilon\}
\end{eqnarray*}
(we put $\mbox{comp}_\varepsilon^q(S,F)=+\infty$ if there is no such algorithm).
It is easily checked that these functions are inverse to each other in the 
following sense: For all $n\in \N_0$ and $\varepsilon > 0$,
$e_n^q(S,F)\le \varepsilon$ if and only if
$\mbox{comp}_{\varepsilon_1}^q(S,F)\le n$ for all $\varepsilon_1 > \varepsilon$.
Hence it suffices to determine one of them. We shall principally choose the first one.

\section{Some General Results}

Let $\emptyset\ne F\subseteq \mathcal{F}(D,K)$ and 
$\emptyset\ne \wt{F}\subseteq \mathcal{F}(\wt{D},\wt{K})$, where $D,\wt{D},K,\wt{K}$ are 
nonempty sets.
Suppose we want to construct an algorithm $A$ on $F$ by the help of some reduction 
to an already known algorithm $\wt{A}$ on $\wt{F}$ in the following form:
For $f\in F$ we construct a function $\wt{f}=\Gamma(f)\in \wt{F}$ and apply $\wt{A}$
to it. When does this indeed give an algorithm on $F$? To clarify the problem, 
note that
 by definition, an  algorithm $A$ on $F$ can only use queries $Q$ on $F$
itself, while in the approach above we use $\wt{Q}_{\Gamma(f)}$ instead, 
where $\wt{Q}$ is a query on $\wt{F}$. 
The way out is to simulate $\wt{Q}_{\Gamma(f)}$ as $B_f$, where  
$B$ is an algorithm without measurement on $F$. 
The following result contains sufficient conditions and is a generalization of 
Lemma 5 of Heinrich (2001a).

Assume that we are
given a mapping $\Gamma:F\to \wt{F}$ of the following type: There are $\kappa,m^*\in\N$ 
and mappings
\begin{eqnarray*}
\eta_j&:& \wt{D}\to D \quad (j=0,\dots,\kappa-1)\\
\beta&:& K\to\Z\big[0,2^{m^*}\big)\\
\rho&:& \wt{D}\times\Z\big[0,2^{m^*}\big)^\kappa \to \wt{K}
\end{eqnarray*}
such that for $f\in F$ and $s\in \wt{D}$
\begin{equation}
\label{F1}
(\Gamma (f))(s)=\rho(s,\beta\circ f\circ\eta_0(s),\dots,\beta\circ f\circ\eta_{\kappa-1}(s)).
\end{equation}
\begin{lemma}
\label{lem:1}
For each quantum query $\wt{Q}$ on $\wt{F}$ and each mapping $\Gamma$ of the above 
form (\ref{F1})
there is a quantum algorithm without measurement $B$ on $F$ such that 
$n_q(B)= 2\kappa$
and for all $f\in F$, $x\in \Z[0,2^{\wt{m}})$,
$$
(\wt{Q}_{\Gamma(f)}\lg x\rs)\lg 0\rs_{m-\wt{m}}=B_f\lg x\rs\lg 0\rs_{m-\wt{m}},
$$
where $\wt{m}=m(\wt{Q})$, $m=m(B)> \wt{m}$ and $\lg 0\rs_{m-\wt{m}}$ stands for the zero 
state of the last $m-\wt{m}$ qubits.
\end{lemma}
\begin{proof} Let 
$$
\wt{Q}=(\wt{m},\wt{m}',\wt{m}'',\wt{Z},\wt{\tau},\wt{\beta}),
$$
and put 
\begin{eqnarray*}
&\kappa_0=\lceil\log \kappa\rceil,\quad      
m=\wt{m}+\kappa_0+\kappa m^*,\quad m'=\wt{m}'+\kappa_0, \quad
m''=m^*,\\
&Z=\wt{Z}\times[0,\kappa),\quad
\tau(i,j)=\eta_j (\wt{\tau}(i)) \quad\mbox{for}\quad (i,j)\in Z,\quad
\end{eqnarray*}
let $\beta$ be as above, and define 
$$
Q=(m,m',m'',Z,\tau,\beta).
$$
We represent 
$$
 H_m=H_{\wt{m}'}\otimes H_{\wt{m}''}\otimes H_{\wt{m}-\wt{m}'-\wt{m}''}\otimes H_{\kappa_0}
 \otimes H_{m^*}^{\otimes \kappa},
$$
a basis state of which will be written as
$$
\lg i\rs\lg x\rs\lg y\rs\lg j\rs\lg z_0\rs\dots\lg z_{\kappa-1}\rs.
$$
Define the permutation operator $P_0$ by
$$
P_0\lg i\rs\lg x\rs\lg y\rs\lg j\rs\lg z_0\rs\dots\lg z_{\kappa-1}\rs
=\lg i\rs\lg j\rs\lg z_0\rs\dots\lg z_{\kappa-1}\rs\lg x\rs\lg y\rs,
$$
another permutation operator
$$
P\lg i\rs\lg j\rs\lg z_0\rs\dots\lg z_{j}\rs\dots\lg z_{\kappa-1}\rs\lg x\rs\lg y\rs
=\lg i\rs\lg j\rs\lg z_{j}\rs\dots\lg z_0\rs\dots\lg z_{\kappa-1}\rs\lg x\rs\lg y\rs,
$$
the following counting operators
\begin{eqnarray*}
C_0\lg i\rs\lg j\rs\dots\lg y\rs&=&\lg i\rs\lg j\oplus \kappa\rs\dots\lg y\rs\\
C\lg i\rs\lg j\rs\dots\lg y\rs&=&\lg i\rs\lg j\oplus 1\rs\dots\lg y\rs,
\end{eqnarray*}
where $\oplus$ is addition modulo $2^{\kappa_0}$, 
and the operator of sign inversion
$$
J\lg i\rs\lg j\rs\lg z_0\rs\dots\lg y\rs
=\lg i\rs\lg j\rs\lg \ominus z_0\rs\dots\lg y\rs,
$$
where $\ominus$ is subtraction modulo $2^{m^*}$ and $\ominus z$ stands for
$0\ominus z$.
Finally, let 
\begin{eqnarray*}
\lefteqn{
T\lg i\rs\lg j\rs\lg z_0\rs\dots\lg z_{\kappa-1}\rs\lg x\rs\lg y\rs}\\
&=&\lg i\rs\lg j\rs\lg z_0\rs\dots\lg z_{\kappa-1}\rs
\lg x\oplus\wt{\beta}\circ\rho(\wt{\tau}(i),z_0,\dots, 
z_{\kappa-1})\rs\lg y\rs
\end{eqnarray*}
if $i\in \wt{Z}$, and 
$$
T\lg i\rs\lg j\rs\lg z_0\rs\dots\lg z_{\kappa-1}\rs\lg x\rs\lg y\rs
=\lg i\rs\lg j\rs\lg z_0\rs\dots\lg z_{\kappa-1}\rs\lg x\rs\lg y\rs
$$
if $i\not\in \wt{Z}$.
We define $B$ by setting for $f\in F$,
$$
B_f=P_0^{-1}C_0^{-1} (CQ_f JP)^\kappa T (PQ_f C^{-1})^\kappa C_0 P_0.
$$
Let us trace the action of $B_f$ on
$$
\lg i\rs\lg x\rs\lg y\rs\lg 0\rs\lg 0\rs\dots\lg 0\rs.
$$
First we assume $i\in \wt{Z}$. The application of $P_0$, followed by $C_0$, gives
$$
\lg i\rs\lg \kappa\ \mbox{mod}\ 2^{\kappa_0}\rs\lg 0\rs\dots\lg 0\rs\lg x\rs\lg y\rs.
$$
The transformation $PQ_f C^{-1}$ leads to
$$
\lg i\rs\lg \kappa-1\rs\lg 0\rs\dots
\lg\beta\circ f\circ\eta_{\kappa-1}\circ\wt{\tau}(i) \rs\lg x\rs\lg y\rs,
$$
and after the remaining $\kappa-1$ applications of $PQ_f C^{-1}$ we get
$$
\lg i\rs\lg 0\rs\lg  \beta\circ f\circ\eta_{0}\circ\wt{\tau}(i)\rs\dots
\lg\beta\circ f\circ\eta_{\kappa-1}\circ\wt{\tau}(i) \rs\lg x\rs\lg y\rs.
$$
 Then the above is mapped by $T$ to
\begin{eqnarray*}
&&\lefteqn{\lg i\rs\lg 0\rs\lg  \beta\circ f\circ\eta_{0}\circ\wt{\tau}(i)\rs\dots
\lg\beta\circ f\circ\eta_{\kappa-1}\circ\wt{\tau}(i) \rs}\\
&&\lg x\oplus\wt{\beta}\circ\rho(\wt{\tau}(i),\beta\circ f\circ\eta_0\circ\wt{\tau}(i),\dots, 
\beta\circ f\circ\eta_{\kappa-1}\circ\wt{\tau}(i))\rs\lg y\rs\\
&=&\lg i\rs\lg 0\rs\lg  \beta\circ f\circ\eta_{0}\circ\wt{\tau}(i)\rs\dots
\lg\beta\circ f\circ\eta_{\kappa-1}\circ\wt{\tau}(i) \rs\\
&&\lg x\oplus \wt{\beta}((\Gamma (f))(\wt{\tau}(i)))\rs\lg y\rs.
\end{eqnarray*} 
The transformation $(CQ_f JP)^\kappa$ produces
$$
\lg i\rs\lg \kappa\ \mbox{mod}\ 2^{\kappa_0}\rs\lg 0\rs\dots\lg 0\rs\lg x\oplus 
\wt{\beta}((\Gamma (f))(\wt{\tau}(i)))\rs\lg y\rs,
$$
and finally $P_0^{-1}C_0^{-1}$ gives
$$
\lg i\rs\lg x\oplus \wt{\beta}((\Gamma (f))(\wt{\tau}(i)))\rs\lg y\rs\lg 0\rs\lg 0\rs
\dots\lg 0\rs
=(\wt{Q}_{\Gamma(f)}\lg i\rs\lg x\rs\lg y\rs)\lg 0\rs_{m-\wt{m}}.
$$
The case $i\not\in \wt{Z}$ is checked analogously.
\end{proof}
\begin{corollary}
\label{cor:1}
Given a mapping $\Gamma:F\to \wt{F}$ as in (\ref{F1}), a normed space
$G$ and a quantum algorithm $\wt{A}$ from $\wt{F}$ to $G$, there is a quantum algorithm 
$A$ from $F$ to $G$ with 
\[
n_q(A)=2\kappa\,n_q(\wt{A})
\]
and for all $f\in F$
$$
A(f)=\wt{A}(\Gamma(f)).
$$
Consequently, if $\wt{S}:\wt{F}\to G$ is any mapping and $S=\wt{S}\circ\Gamma$, 
then for each $n\in \N_0$
$$
e_{2\kappa n}^q(S,F)\le e_n^q(\wt{S},\wt{F}).
$$
\end {corollary} 
The proof is literally the same as that of Corollary 1 in Heinrich (2001a). We omit it here.
\begin{lemma}
\label{lem:2}
Let $D,K$ and $F\subseteq\mathcal{F}(D,K)$ be nonempty sets, let $k\in\N_0$
and let $S_l:F\to \R$ $(l=0,\dots,k)$ be mappings. Define $S:F\to \R$ by
$S(f)=\sum_{l=0}^k S_l(f)\quad(f\in F)$. Let $\theta_0,\dots,\theta_k\ge 0$,
$n_0,\dots,n_k\in\N_0$ and put $n=\sum_{l=0}^k n_l$.   Then
$$
e_n^q(S,F,\sum_{l=0}^k\theta_l)\le\sum_{l=0}^k e_{n_l}^q(S_l,F,\theta_l).
$$
\end{lemma}

\begin{proof}
Let $\delta>0$ and let $A_l$ be a quantum algorithm from $F$ to $\R$ with 
$n_q(A_l)\le n_l$ and 
$$
e(S_l,A_l,F,\theta_l)\le e_{n_l}^q(S_l,F,\theta_l)+\delta.
$$
Let $A=\sum_{l=0}^k A_l$ be the composed algorithm (in the sense of section
2 of Heinrich, 2001a). Then
\begin{equation}
\label{C3}
n_q(A)=\sum_{l=0}^k n_q(A_l)\le\sum_{l=0}^k n_l.
\end{equation}
Fix an $f\in F$ and let $(\zeta_{l,f})_{l=0}^k$ be independent 
random variables with distribution $A_l(f)$. It follows that
with probability at least $1-\theta_l$,
$$
|S_l(f)-\zeta_{l,f}|\le e_{n_l}^q(S_l,F,\theta_l)+\delta.
$$
Setting  
$$
\zeta_f=\sum_{l=0}^k\zeta_{l,f},
$$
we infer from  Lemma 2 of Heinrich (2001a) that $\zeta_f$ has distribution
$A(f)$. Consequently,
$$
|S(f)-\zeta_f|=\Big|\sum_{l=0}^k (S_l(f)-\zeta_{l,f})\Big|
\le \sum_{l=0}^k (e_{n_l}^q(S_l,F,\theta_l)+\delta)
$$
with probability at least 
$$
1-\sum_{l=0}^k \theta_l.
$$
This gives
$$
e(S,A,f,\sum_{l=0}^k \theta_l)\le \sum_{l=0}^k e_{n_l}^q(S_l,F,\theta_l)\,+\,(k+1)\delta
$$
for all $f\in F$, and the desired result follows from (\ref{C3}).
\end{proof}

\begin{corollary}
\label{cor:3}
Let $D,K$,  $F\subseteq\mathcal{F}(D,K)$, $k\in\N_0$
and $S,S_l:F\to \R$ $(l=0,\dots,k)$ be as in Lemma \ref{lem:2}. 
 Assume $\nu_0,\dots,\nu_k\in \N$ satisfy
$$
\sum_{l=0}^k e^{-\nu_l/8}\le \frac{1}{4}.
$$ 
Let $n_0,\dots,n_k\in\N_0$ and put
$n=\sum_{l=0}^k \nu_l n_l$. Then
$$
e_n^q(S,F)\le\sum_{l=0}^k e_{n_l}^q(S_l,F).
$$
\end{corollary}

This is an obvious consequence of Lemma \ref{lem:2} above and of Lemma 3 
in Heinrich (2001a), which can be restated in the following form:
\begin{lemma}
\label{lem:3}
Let $S$ be any mapping from $F\subseteq\mathcal{F}(D,K)$ to $\R$. 
Then for each $n,\nu\in\N$,
$$
e_{\nu n}^q(S,F,e^{-\nu/8})\le e_n^q(S,F).
$$
\end{lemma}

\section{Summation}
This section provides the prerequisites from summation needed later for the study
of integration.
 For $N\in\N$ and  $1\le p<\infty$, 
let $L_p^N$ denote the space of all functions
$f:\Z[0,N)\to \R$, equipped with the norm 
$$
\|f\|_{L_p^N}=\left(\frac{1}{N}\sum_{i=0}^{N-1}|f(i)|^p \right)^{1/p}.
$$
Define $S_N:L_p^N\to\R$ by 
$$
S_N f=\frac{1}{N}\sum_{i=0}^{N-1}f(i), 
$$  and let
$$
\mathcal{B}(L_p^N):=\{f\in L_p^N \,|\, \|f\|_{L_p^N}\le 1\}
$$
be the unit ball of $L_p^N$.

We need the following results about summation, where (\ref{K1}) and (\ref{K2}) are from
Heinrich (2001a), Theorems 1 and 2, and (\ref{K3}) is 
from Heinrich and Novak (2001b), Corollary 2.
\begin{proposition}
\label{pro:1} Let $1\le p<\infty$.
There are constants $c_1,c_2,c_3>0$ 
such that for all $n,N\in\N$ with $n\le c_1 N$,
\begin{equation}
\label{K1}
c_2 n^{-1}\le e_n^q(S_N,\mathcal{B}(L_p^N))\le c_3 n^{-1}\quad\mbox{if}\quad 2<p<\infty,
\end{equation}
\begin{equation}
\label{K2}
c_2 n^{-1}\le e_n^q(S_N,\mathcal{B}(L_2^N))\le c_3 n^{-1}\log^{3/2}n\log\log n,
\end{equation}
and 
\begin{eqnarray}
\lefteqn{
c_2 \min(n^{-2(1-1/p)},n^{-2/p}N^{2/p-1})\le e_n^q(S_N,\mathcal{B}(L_p^N))}\nonumber\\
&&
\le c_3\min(n^{-2(1-1/p)},n^{-2/p}N^{2/p-1})\max(\log(n/\sqrt{N}),1)^{2/p-1}.\label{K3}
\end{eqnarray}
if $1\le p<2$.
\end{proposition}
\noindent {\bf Remark.} 
We often use the same symbol $c,c_1,\dots$ for possibly different
positive constants (also when they appear in a sequence of relations).
These constants are either absolute or may depend only on $p,r,d$ -- 
in all lemmas and the theorem this is precisely described anyway by the order of the
quantifiers.\\
\\
In the case $p=2$ we will not use Proposition \ref{pro:1}
alone -- that would give just a logarithmic factor instead of the iterated logarithm
of Theorem \ref{theo:1} below. 
In the region where $n$ 
is close to $N$ we use a refinement which can be obtained on the basis of the results in 
Heinrich and Novak (2001b). We introduce
for $M\in\N$
$$
S_{N,M}f=\frac{1}{N}\sum_{i\in\Z[0,N],\,|f(i)|<M}f(i)
$$
and
$$
S'_{N,M}f=S_Nf-S_{N,M}f=\frac{1}{N}\sum_{i\in\Z[0,N],\,|f(i)|\ge M}f(i).
$$
Let us first recall the case $p=2$ of Corollary 3 of Heinrich and Novak (2001b), 
which we will use here:
\begin{lemma}\label{lem:4}
There is a constant $c>0$ such that for all $n,M,N\in\N$,
$$
e_n^q(S'_{N,M},\mathcal{B}(L_2^N))=0
$$
whenever
$$
M\ge cNn^{-1}\max(\log(n/\sqrt{N}),1).
$$
\end{lemma}
The next result can be shown by repeating the 
respective part of the proof of Theorem 1 of Heinrich (2001a) (for the analogous 
$p<2$ case see also the proof of Proposition 2 in Heinrich and Novak, 2001b). 
\begin{lemma}
\label{lem:5}
There is a constant $c>0$ such that for all $k,n,N\in\N,\,\,k>1$
$$
e_n^q(S_{N,2^k},\mathcal{B}(L_2^N))\le c\left(n^{-1}k^{3/2}\log k 
+2^k n^{-2}(k\log k)^2\right).
$$
\end{lemma}
From these we can derive the following estimate:
\begin{lemma}\label{lem:6}
There is a constant $c>0$ such that for all $n,N\in\N$
with $n \le N$,
$$
e_n^q(S_N,\mathcal{B}(L_2^N))\le cn^{-1}\l(n,N)^{3/2}\log\l(n,N),
$$
where 
$$
\l(n,N)=\log(N/n)+\log\log (n+1)+2.             
$$
\end{lemma}
\noindent {\bf Remark.} Observe that Lemma \ref{lem:6} gives an improvement over
Proposition \ref{pro:1} only for $n$ close to $N$.
\begin{proof} It suffices to prove the statement for 
\begin{equation}
\label{M7}
N\le n^{3/2},
\end{equation}
the other case follows directly from (\ref{K2}).
Let $c_0$ denote the constant from Lemma \ref{lem:4} and 
let $k$ be the smallest natural number with $k\ge 2$ and
\begin{equation}
\label{I6}
c_0 Nn^{-1}\max(\log(n/\sqrt{N}),1)\le 2^k.
\end{equation}
By Lemma \ref{lem:4},
\begin{equation}
\label{I9}
e_n^q(S'_{N,2^k},\mathcal{B}(L_2^N))=0.
\end{equation}
This together with Lemma \ref{lem:5} and Corollary \ref{cor:3} gives
\begin{equation}
\label{J1}
e_{c_1 n}^q(S_N,\mathcal{B}(L_2^N))\le cn^{-1}k^{3/2}\log k +c\,2^k n^{-2}(k\log k)^2, 
\end{equation}
with a certain constant $c_1\in \N$.
It follows from (\ref{I6}) that
\begin{equation}
\label{M1}
2^{k-1}\le \max\big(c_0 Nn^{-1}\max(\log(n/\sqrt{N}),1),\, 2\big),
\end{equation}
which, in turn, implies
\begin{equation}
\label{M2}
2^k\le cNn^{-1}\log (n+1),
\end{equation}
\begin{equation}
\label{M4}
k\le c(\log(N/n)+\log\log (n+1)+1)=c\l(n,N),
\end{equation}
and thus 
\begin{equation}
\label{M6}
k\le c\log(N+1).
\end{equation}
From (\ref{M4}) and $\l(n,N)\ge 2$ we conclude 
\begin{equation}
\label{N2A}
\log k\le c\log\l(n,N),
\end{equation}
while (\ref{M6}) gives 
\begin{equation}
\label{M5}
\log k\le c(\log\log(N+1)+1).
\end{equation}
From (\ref{M2}), (\ref{M6}), (\ref{M5}), and (\ref{M7}), we infer 
\begin{eqnarray*}
2^k n^{-1} k^{1/2}(\log k)^2 
&\le& cNn^{-2}\log(n+1)(\log N)^{1/2}(\log\log (N+1)+1)^2 \\
&\le& cN^{-1/3}(\log (N+1))^{3/2}(\log\log (N+1)+1)^2 \le c.
\end{eqnarray*}
Consequently 
$$
2^k n^{-2}(k\log k)^2 \le cn^{-1}k^{3/2}\log k,
$$
and hence, by (\ref{J1}), (\ref{M4}), and (\ref{N2A}),
$$
e_{c_1 n}^q(S_N,\mathcal{B}(L_2^N))\le cn^{-1}k^{3/2}\log k
\le cn^{-1}\l(n,N)^{3/2}\log\l(n,N).
$$
\end{proof}

\section{Integration}
This section contains the main result. Let $D=[0,1]^d$ and 
let $C(D)$ be the space of continuous functions on $D$, equipped with the supremum norm.
For 
$1\le p <\infty$, let $L_p(D)$
be the space of $p$-integrable functions, endowed with the usual norm
$$\|f\|_{L_p(D)}=\left(\int_D|f(t)|^pdt\right)^{1/p}.$$  
The Sobolev space
$W_p^r(D)$ consists of all functions $f\in L_p(D)$ such that for all 
$\a=(\a_1,\dots,\a_d)\in N_0^d$
with $|\a|:=\sum_{j=1}^d\a_j\le r$, 
the generalized partial derivative $\partial^\a f$ belongs to $L_p(D)$. The norm 
on $W_p^r(D)$ is defined as 
$$\|f\|_{W_p^r(D)}=\left(\sum_{|\a|\le r}\|\partial^\a f\|^p_{L_p(D)}\right)^{1/p}.$$
We shall assume that $r/d>1/p$, which, by the Sobolev embedding theorem (see
Adams, 1975, or Triebel, 1995), implies that functions from $W_p^r(D)$ are continuous
on $D$, and hence function values are well defined. Let $\mathcal{B}(W_p^r(D))$ be
the unit ball of $W_p^r(D)$ and let $I_d:W_p^r(D)\to \R$ be the integration operator
$$
I_d(f)=\int_Df(t)dt.
$$

\begin{theorem}
\label{theo:1}
Let $r,d\in\N$, $1\le p <\infty$ and assume $r/d>1/p$. 
There are constants $c_1,c_2>0$
such that for all $n\in\N$ with $n>4$
\begin{eqnarray*}
c_1 n^{-r/d-1}&\le& e_n^q(I_d,\mathcal{B}(W_p^r(D)))\le c_2 n^{-r/d-1}
\quad\mbox{if}\quad 2<p<\infty,\\
c_1 n^{-r/d-1}&\le& e_n^q(I_d,\mathcal{B}(W_2^r(D)))\le c_2 n^{-r/d-1}\lambda_0(n),\\
c_1 n^{-r/d-1}&\le& e_n^q(I_d,\mathcal{B}(W_p^r(D)))\le c_2 n^{-r/d-1}(\log n)^{2/p-1}
\quad\mbox{if}\quad 1\le p<2.
\end{eqnarray*}
The function $\lambda_0$ denotes an iterated-logarithmic factor: 
$$
\lambda_0(n)=(\log\log n)^{3/2}\log\log\log n.
$$
\end{theorem}
\begin{proof} First we prepare the needed tools for the upper bound proof.
For $l\in\N_{0}$ let 
\begin{eqnarray*}
D= \bigcup_{i=0}^{2^{dl}-1} D_{li}
\end{eqnarray*}
be the partition of $D$ into $2^{dl}$  congruent cubes of disjoint interior.
Let $s_{li}$ denote the point in $D_{li}$ with the smallest Euclidean norm.
We introduce the following extension operator  
$$
E_{li}: \mathcal{F}(D,\R)\to \mathcal{F}(D,\R)
$$
by setting 
$$
(E_{li}f)(s)=f(s_{li}+2^{-l}s)
$$
for $f\in\mathcal{F}(D,\R)$ and $s\in D$.
Now let $J$ be any quadrature rule on $C(D)$,  
$$
Jf=\sum_{j=0}^{\kappa-1} a_j f(t_j)\quad(f\in C(D))
$$
with $a_j\in\R$ and $t_j\in D$,  which is exact on $\mathcal{P}_{r-1}(D)$,
that is, 
\begin{equation}
\label{A1}
Jf=I_d f \quad \mbox{for all}\quad f\in\mathcal{P}_{r-1}(D),
\end{equation}
 where
$\mathcal{P}_{r-1}(D)$ denotes the space of
polynomials on $D$ of degree not exceeding $r-1$. (For example, for $d=1$ 
one can take the Newton-Cotes formulas of appropriate degree and
for $d>1$ their tensor products.) Since $r>d/p$, we have, by the Sobolev embedding
theorem (see Adams, 1975, or Triebel, 1995), 
$W_p^r (D)\subset C(D)$ and there is a constant $c>0$ such that for
each $f\in W_p^r (D)$
\begin{equation}
\label{E1}
\|f\|_{C(D)}\le c\|f\|_{W_p^r (D)}.
\end{equation}
Consequently,
\begin{equation}
\label{A2}
|Jf|\le\sum_{j=0}^{\kappa-1} |a_j||f(t_j)|\le \sum_{j=0}^{\kappa-1} |a_j|\|f\|_{C(D)}\le c\|f\|_{W_p^r (D)}.
\end{equation}
For $f\in W_p^r (D)$ we denote 
$$
|f|_{r,p,D}=\left(\sum_{|\alpha|=r}\int_D |\partial^\alpha f(t)|^p\,dt\right)^{1/p}.
$$
According to Theorem 3.1.1 in Ciarlet (1978), there is a constant $c>0$ such that 
for all $f\in W_p^r (D)$
\begin{equation}
\label{B1}
\inf_{g\in\mathcal{P}_{r-1}(D)} \|f-g\|_{W_p^r (D)}
\le c|f|_{r,p,D}.
\end{equation}
We conclude from (\ref{A1}), (\ref{A2}) and (\ref{B1}),
\begin{eqnarray}
|I_df-Jf|&\le&\inf_{g\in\mathcal{P}_{r-1}(D)}|I_d (f-g)-J(f-g)|\nonumber\\ 
&\le& c\inf_{g\in\mathcal{P}_{r-1}(D)} \|f-g\|_{W_p^r (D)}
\le c |f|_{r,p,D}\label{A4}.
\end{eqnarray}
Now define for $l\in\N_0$  
$$
J_l f = 2^{-dl}\sum_{i=0}^{2^{dl}-1} J(E_{li}f)
=2^{-dl}\sum_{i=0}^{2^{dl}-1}\sum_{j=0}^{\kappa-1} a_j f(s_{li}+2^{-l}t_j),
$$
which is the composed quadrature obtained by scaling $J$ to the subcubes $D_{li}$.
Then we have 
for $f\in W_p^r (D)$ 
\begin{eqnarray}
|I_df-J_lf|&=&|I_df-2^{-dl}\sum_{i=0}^{2^{dl}-1}J(E_{li}f)|\nonumber\\
&\le&2^{-dl}\sum_{i=0}^{2^{dl}-1}|I_d(E_{li}f)-J(E_{li}f)|\nonumber\\
&\le& c\,2^{-dl}\sum_{i=0}^{2^{dl}-1}|E_{li}f|_{r,p,D}\nonumber\\
&\le& c\left(2^{-dl}\sum_{i=0}^{2^{dl}-1}|E_{li}f|_{r,p,D}^p\right)^{1/p}\nonumber
\end{eqnarray}
and
\begin{eqnarray}
2^{-dl}\sum_{i=0}^{2^{dl}-1}|E_{li}f|_{r,p,D}^p
&=&2^{-dl}\sum_{i=0}^{2^{dl}-1}\sum_{|\alpha|=r}
\int_D |\partial^\alpha f(s_{li}+2^{-l}t)|^p\,dt\nonumber\\
&=&2^{-prl}\sum_{i=0}^{2^{dl}-1}\sum_{|\alpha|=r}
\int_{D_{li}} |\partial^\alpha f(t)|^p\,dt\nonumber\\
&=&\,2^{-prl}|f|_{r,p,D}^p\le \,2^{-prl}\|f\|_{W_p^r (D)}^p.\label{A5}
\end{eqnarray}
It follows that
\begin{equation}
|I_df-J_lf|\le c\,2^{-rl}|f|_{r,p,D}\le c\,2^{-rl}\|f\|_{W_p^r (D)}.\label{B2}
\end{equation}

Let us now describe the main idea: First we approximate $I_df$ by the quadrature 
$J_kf$ for some $k$, giving the desired
precision, but having a number
of nodes much larger than $n$. This $J_k$, in turn, will be split into the sum of
a single quadrature $J_{k_0}$, with number of nodes of the order $n$, which we compute
classically, and a hierarchy of quadratures (more precisely, differences of quadratures)
$J_l'$ $(l=k_0,\dots, k-1)$. It will be shown that the computation of the $J'_l f$ reduces to the 
computation of the mean of sequences with well-bounded $L_p^{N_l}$-norms for suitable
$N_l$. This enables us to apply the results of section 4 and
approximate the means by quantum algorithms. In the sequel we give the
formal details, the proper balancing of parameters and the proof of the error estimates.

Define
\begin{eqnarray}
J'f:=(J_1-J_0)f
&=&2^{-d}\sum_{i=0}^{2^{d}-1}\sum_{j=0}^{\kappa-1} a_j f(s_{1,i}+2^{-1}t_j)
-\sum_{j=0}^{\kappa-1} a_j f(t_j)\nonumber\\
&:=&\sum_{j=0}^{\kappa'-1} a_j' f(t_j'),\label{A6}
\end{eqnarray}
where 
\begin{equation}
\label{E2}
\kappa'\le\kappa (2^d+1).
\end{equation}
For $l\in\N_0$, set 
\begin{eqnarray}
J'_{li}f&=&J'(E_{li}f)=\sum_{j=0}^{\kappa'-1} a_j' f(s_{li}+2^{-l}t_j')\label{B5},\\
J'_{l} &=& 2^{-dl}\sum_{i=0}^{2^{dl}-1} J'_{li}.\label{B6}
\end{eqnarray}
It is easily checked that
$$
J_{l+1}f=2^{-dl}\sum_{i=0}^{2^{dl}-1}J_1(E_{li}f).
$$
and hence 
\begin{eqnarray}
J_{l+1}f-J_lf&=&2^{-dl}\sum_{i=0}^{2^{dl}-1}(J_1(E_{li}f)-J_0(E_{li}f))\nonumber\\
&=&2^{-dl}\sum_{i=0}^{2^{dl}-1}J'_{li}f=J'_lf.\label{B4}
\end{eqnarray}
Using (\ref{B2}) and (\ref{A5}), we get 
\begin{eqnarray}
2^{-dl}\sum_{i=0}^{2^{dl}-1}|J'_{li}f|^p&\le&
2^{-dl}\sum_{i=0}^{2^{dl}-1}|J_1(E_{li}f)-J_0(E_{li}f)|^p\nonumber\\
&\le&2^{-dl}\sum_{i=0}^{2^{dl}-1}(|(I_d-J_1)(E_{li}f)|+|(I_d-J_0)(E_{li})f)|)^p\nonumber\\
&\le&c\,2^{-dl}\sum_{i=0}^{2^{dl}-1}|E_{li}f|_{r,p,D}^p
\le c\,2^{-prl}\|f\|_{W_p^r (D)}^p.\label{B3}
\end{eqnarray}
Now we derive the upper bounds.
Clearly, it suffices to prove them for 
\begin{equation}
\label{L1}
n\ge \max(\kappa,5).
\end{equation}
 Let  
\begin{equation}
\label{D1}
k_0=\lfloor d^{-1}\log(n/\kappa)\rfloor.
\end{equation}
By the above, we have $k_0\ge 0$. Furthermore, let 
\begin{equation}
\label{H9}
k=\lceil (1+d/r)k_0 \rceil,
\end{equation}
hence $k> k_0$. By (\ref{B4})
\begin{equation}\label{C1}
J_k=J_{k_0}+\sum_{l=k_0}^{k-1}J'_l.
\end{equation}
For
\begin{equation}
\label{O1}
k_0\le l<k
\end{equation}
put $N_l=2^{dl}$. We shall define mappings 
$\Gamma_l:\mathcal{B}(W_p^r(D))\to L_p^{N_l}$ in order to apply Lemma \ref{lem:1}.
For this purpose we fix an $m^*\in \N$ with 
\begin{equation}
\label{G1}
2^{-m^*/2}\le k^{-1}2^{-rk}
\end{equation}
and
\begin{equation}
\label{G2}
2^{m^*/2-1}\ge c,
\end{equation}
where $c$ is the constant from (\ref{E1}).
Hence,
\begin{equation}
\label{E3}
\|f\|_{C(D)}\le 2^{m^*/2-1}\quad \mbox{for}\quad f\in\mathcal{B}(W_p^r(D)).
\end{equation}
Define $\eta_{lj}:\Z[0,N_l)\to D\quad (j=0,\dots.\kappa'-1)$ by
$$
\eta_{lj}(i)=s_{li}+2^{-l}t_j'\quad (i\in\Z[0,N_l)),
$$
and $\b:\R\to\Z[0,2^{m^*})$ for $z\in\R$ by
\begin{equation}\label{N1}
\b(z)=
\left\{\begin{array}{lll}
   0& \mbox{if} \quad z <-2^{m^*/2-1} \\
   \lfloor 2^{m^*/2}(z+2^{m^*/2-1})\rfloor       & \mbox{if} \quad  
   -2^{m^*/2-1}\le z <2^{m^*/2-1}\\
   2^{m^*}-1& \mbox{if} \quad z\ge 2^{m^*/2-1}. 
   \end{array}
   \right.
\end{equation}
Furthermore, let $\gamma:\Z[0,2^{m^*})\to\R$ be defined for $y\in\Z[0,2^{m^*})$ as 
\begin{equation}
\label{N2}
\g(y)=2^{-m^*/2}y-2^{m^*/2-1}.
\end{equation}
It follows that for $-2^{m^*/2-1}\le z\le 2^{m^*/2-1}$,
\begin{equation}
\label{E4}
\g(\b(z))\le z\le \g(\b(z))+2^{-m^*/2}.
\end{equation}
Next let $\rho:\Z[0,2^{m^*})^{\kappa'}\to\R$ be given by
$$
\rho(y_0,\dots,y_{\kappa'-1})=\sum_{j=0}^{\kappa'-1}a_j'\g(y_j).
$$
Finally, we set
$$
\Gamma_l(f)(i)=\rho((\b\circ f\circ\eta_{lj}(i))_{j=0}^{\kappa'-1}).
$$
for $f\in \mathcal{B}(W_p^r(D))$.
We have 
$$
\Gamma_l(f)(i)=\sum_{j=0}^{\kappa'-1}a_j'\g(\b(f(s_{li}+2^{-l}t_j'))),
$$
hence, by (\ref{B5}), (\ref{E3}) and (\ref{E4}),
\begin{eqnarray}
\label{H1}
|J'_{li}f-\Gamma_l(f)(i)|&\le& \sum_{j=0}^{\kappa'-1}|a_j'||f(s_{li}+2^{-l}t_j')
-\g(\b(f(s_{li}+2^{-l}t_j')))|\nonumber\\
&\le& 2^{-m^*/2}\sum_{j=0}^{\kappa'-1}|a'_j|\le c\,2^{-m^*/2}\le ck^{-1}2^{-rk},
\end{eqnarray}
and therefore, by (\ref{B6}),  for all $f\in \mathcal{B}(W_p^r(D))$,
\begin{equation}
\label{E5}
|J'_lf-S_{N_l}\Gamma_l(f)|\le 2^{-dl}\sum_{i=0}^{2^{dl}-1}|J'_{li}f-\Gamma_l(f)(i)|
\le ck^{-1}2^{-rk}. 
\end{equation}
Moreover, by (\ref{B3}),  (\ref{H1}), and (\ref{O1}),
\begin{eqnarray*}
\|\Gamma_l(f)\|_{L_p^{N_l}}&\le& \|(J_{li}'f)_{i=0}^{N_l-1}\|_{L_p^{N_l}}
+\|\Gamma_l(f)-(J_{li}'f)_{i=0}^{N_l-1}\|_{L_p^{N_l}}\\
&\le& \|(J_{li}'f)_{i=0}^{N_l-1}\|_{L_p^{N_l}}
+\|\Gamma_l(f)-(J_{li}'f)_{i=0}^{N_l-1}\|_{L_\infty^{N_l}}\\
&\le& c\,2^{-rl}.
\end{eqnarray*}
Consequently, 
\begin{equation}
\label{C4}
\Gamma_l (\mathcal{B}(W_p^r(D)))\subseteq c\,2^{-rl}\mathcal{B}(L_p^{N_l}).
\end{equation}
By (\ref{D1}), $\kappa\,2^{dk_0}\le n$, hence
\begin{equation}
\label{H3}
e_{n}^q(J_{k_0},\mathcal{B}(W_p^r(D)),0)=0
\end{equation}
(this just means that with $\kappa\,2^{dk_0}$ queries
we can compute $J_{k_0}$, the mean of 
$\kappa\,2^{dk_0}$ numbers, classically, or, more precisely, up to any precision by  
simulating the classical computation on a suitable number of qubits).
By assumption, $r/d>1/p$ and $p\ge 1$. Hence 
$$
r>\frac{d}{p}\ge\left(\frac{2}{p}-1\right)d.
$$
Now fix any $\delta$ with 
\begin{equation}
\label{N5}
0<\delta<\min\left(r,\frac{p}{2}\left(r-\left(\frac{2}{p}-1\right)d\right)\right),
\end{equation}
and put for $l=k_0,\dots,k-1$
\begin{equation}
\label{I1}
 n_l=\left\lceil 2^{dk_0-\delta(l-k_0)}\right\rceil,
\end{equation}
\begin{equation}
\label{I1A} 
 \nu_l=\left\lceil 8(2\ln(l-k_0+1)+ \ln 8)\right\rceil. 
\end{equation}
It follows from (\ref{I1A}) that 
\begin{equation}
\label{I2}
\sum_{l=k_0}^{k-1}e^{-\nu_l/8}\le \frac{1}{8}\sum_{l=k_0}^{k-1}(l-k_0+1)^{-2}<\frac{1}{4}.
\end{equation}
Put
\begin{equation}
\label{C7}
\wt{n}=n+2\kappa'\sum_{l=k_0}^{k-1}\nu_l n_l.
\end{equation}
By (\ref{I1}), (\ref{I1A}), and (\ref{H9}),
\begin{eqnarray}
\label{I3}
\wt{n}&\le& n+2\kappa'\sum_{l=k_0}^{k-1} 
\left\lceil 8(2\ln(l-k_0+1)+ \ln 8)\right\rceil 
\left\lceil2^{dk_0-\delta(l-k_0)}\right\rceil\nonumber\\
&\le& c\,2^{dk_0} \le cn.
\end{eqnarray}
From (\ref{B2}) above and Lemma 6(i) of Heinrich (2001a),
\begin{equation}
\label{H2}
e_{\wt{n}}^q(I_d,\mathcal{B}(W_p^r(D)))\le
c\,2^{-rk}+e_{\wt{n}}^q(J_k,\mathcal{B}(W_p^r(D))).
\end{equation}
By Lemma \ref{lem:2} and (\ref{H3}),
\begin{eqnarray}
\label{H4}
\lefteqn{e_{\wt{n}}^q(J_k,\mathcal{B}(W_p^r(D)))}\nonumber\\
&\le&
e_n^q(J_{k_0},\mathcal{B}(W_p^r(D)),0)+e_{\wt{n}-n}^q(J_k-J_{k_0},\mathcal{B}(W_p^r(D)))
\nonumber\\
&=& e_{\wt{n}-n}^q(J_k-J_{k_0},\mathcal{B}(W_p^r(D))).
\end{eqnarray}
Using (\ref{C1}), (\ref{C7}), Corollary \ref{cor:3}  and  (\ref{I2}), we get
\begin{eqnarray}
\label{H5}
e_{\wt{n}-n}^q(J_k-J_{k_0},\mathcal{B}(W_p^r(D)))&=&
e_{2\kappa'\sum_{l=k_0}^{k-1}\nu_l n_l}^q\left(\sum_{l=k_0}^{k-1}J'_l,
\mathcal{B}(W_p^r(D))\right)\nonumber\\
&\le&\sum_{l=k_0}^{k-1} e_{2\kappa'n_l}^q(J'_l,\mathcal{B}(W_p^r(D))).
\end{eqnarray}
From  (\ref{E5}) above and Lemma 6(i) of Heinrich (2001a) we conclude
\begin{equation}
\label{H6}
e_{2\kappa'n_l}^q(J'_l,\mathcal{B}(W_p^r(D)))\le
ck^{-1}2^{-rk}+e_{2\kappa'n_l}^q(S_{N_l}\Gamma_l,\mathcal{B}(W_p^r(D))).
\end{equation}
Corollary \ref{cor:1}, relation (\ref{C4}) above and Lemma 6(iii) of Heinrich (2001a) give
\begin{eqnarray}
\label{H7}
e_{2\kappa'n_l}^q(S_{N_l}\Gamma_l,\mathcal{B}(W_p^r(D)))
&\le& e_{n_l}^q(S_{N_l},c\,2^{-rl}\mathcal{B}(L_p^{N_l}))\nonumber\\
&=& c\,2^{-rl}e_{n_l}^q(S_{N_l},\mathcal{B}(L_p^{N_l})).
\end{eqnarray}
Joining (\ref{H2})--(\ref{H7}),
we conclude
\begin{equation}
\label{H8}
e_{\wt{n}}^q(I_d,\mathcal{B}(W_p^r(D)))
\le c\,2^{-rk}+c\sum_{l=k_0}^{k-1} 2^{-rl}e_{n_l}^q(S_{N_l},\mathcal{B}(L_p^{N_l})).
\end{equation}
Now we prove the upper bound in the case $2<p<\infty$. 
Relation (\ref{H8}), Proposition \ref{pro:1}, (\ref{I1}), and (\ref{H9})  give
\begin{eqnarray*}
e_{\wt{n}}^q(I_d,\mathcal{B}(W_p^r(D)))
&\le& c\,2^{-rk}+c\sum_{l=k_0}^{k-1} 2^{-rl}n_l^{-1}\\
&\le& c\,2^{-rk}+c\,2^{-(r+d)k_0}\sum_{l=k_0}^{k-1} 2^{-(r-\delta)(l-k_0)}\\
&\le& c\,2^{-(r+d)k_0}\le cn^{-r/d-1}.
\end{eqnarray*}
Next we consider the case $1\le p <2$. Observe that, by (\ref{N5}),
\begin{equation}
\label{I5}
\frac{2}{p}\delta<r-\left(\frac{2}{p}-1\right)d.
\end{equation}
It follows from (\ref{H8}), 
Proposition \ref{pro:1}, (\ref{I5}), (\ref{H9}), and (\ref{D1}) that
\begin{eqnarray*}
\lefteqn{e_{\wt{n}}^q(I_d,\mathcal{B}(W_p^r(D)))}\\
&\le& c\,2^{-rk}+c\sum_{l=k_0}^{k-1} 2^{-rl}
n_l^{-2/p}N_l^{2/p-1}\max(\log(n_l/\sqrt{N_l}),1)^{2/p-1}\\
&\le& c\,2^{-rk}+c\sum_{l=k_0}^{k-1} 
2^{-rl-\frac{2}{p}dk_0+\frac{2}{p}\delta(l-k_0)+(\frac{2}{p}-1)dl}
(k_0+1)^{\frac{2}{p}-1}\\
&\le& c\,2^{-rk}+c\,2^{-(r+d)k_0}(k_0+1)^{\frac{2}{p}-1}
\sum_{l=k_0}^{k-1}2^{(-r+(\frac{2}{p}-1)d+\frac{2}{p}\delta)(l-k_0)}\\
&\le& c\,2^{-(r+d)k_0}(k_0+1)^{\frac{2}{p}-1}\le cn^{-r/d-1}(\log n)^{\frac{2}{p}-1}.
\end{eqnarray*}
Finally we consider the case $p=2$. From (\ref{H8}) and Lemma \ref{lem:6}
we get
\begin{eqnarray*}
\lefteqn{e_{\wt{n}}^q(I_d,\mathcal{B}(W_p^r(D)))}\\
&\le& c\,2^{-rk}+c\sum_{l=k_0}^{k-1} 2^{-rl}
n_l^{-1}\l(n_l,N_l)^{3/2}\log\l(n_l,N_l)\\
&\le& c\,2^{-rk}+c\,2^{-(r+d)k_0}\times\\
&&\sum_{l=k_0}^{k-1} 
2^{-(r-\delta)(l-k_0)}(l-k_0+\log\log n+1)^{3/2}
\log(l-k_0+\log\log n+1)\\
&\le& c\,2^{-(r+d)k_0}(\log\log n)^{3/2}\log\log\log n\le cn^{-r/d-1}\l_0(n).
\end{eqnarray*}
To conclude the proof of the upper bounds in all three cases, 
we use (\ref{I3}) and scale $n$.

Now we turn to the lower bounds.
Since $\mathcal{B}(W_p^r(D))\subset\mathcal{B}(W_q^r(D))$ for $p>q$, it suffices
to consider the case $2<p<\infty$. Fix such a $p$.
Let $\psi$ be a $C^\infty$ function on $\R^d$ with 
$$
{\rm supp}\, \psi\subset (0,1)^d, \quad \sigma_1:=I_d\psi>0,
$$
and denote $\|\psi\|_{W_p^r(D)}=\sigma_2$.
Let $n\in \N$, $k=\lceil d^{-1}(\log(n/c_1)+1)\rceil$, where $c_1$ is the constant from 
Proposition \ref{pro:1}, which can be assumed to satisfy $0<c_1\le 1$, and put $N=2^{dk}$.
 It follows that 
\begin{equation}
\label{N6}
c_12^{-d}2^{dk}\le 2n\le c_1 2^{dk} = c_1 N.
\end{equation}
Set
$$
\psi_i(t)=\psi(2^k(t-s_i))\quad (i=0,\dots,N-1),
$$
with the $s_i$ as in the beginning of the proof. We have
\begin{equation}
\label{G7}
I_d\psi_i=2^{-dk}I_d\psi=\sigma_1 2^{-dk}=\sigma_1 N^{-1}
\end{equation}
and
$$
\|\psi_i\|_{W_p^r(D)}\le 2^{(r-d/p)k}\|\psi\|_{W_p^r(D)}=\sigma_2 2^{(r-d/p)k}.
$$
Consequently, taking into account the disjointness of the supports
of the $\psi_i$, for all $a_i\in\R\quad (i=0,\dots,N-1)$,
\begin{equation}
\label{N7}
\Big\|\sum_{i=0}^{N-1}a_i\psi_i\Big\|_{W_p^r(D)}^p
=\sum_{i=0}^{N-1}|a_i|^p\|\psi_i\|_{W_p^r(D)}^p
\le \sigma_2^p 2^{prk}\big\|(a_i)_{i=0}^{N-1}\big\|_{L_p^N}^p. 
\end{equation}
Fix any $m^*\in\N$ with 
\begin{equation}
\label{N8}
m^*/2-1\ge dk/p.
\end{equation}
Let $\b:\R\to\Z[0,2^{m^*})$ and $\gamma:\Z[0,2^{m^*})\to\R$ be defined as in
(\ref{N1}) and (\ref{N2}). For $f\in \mathcal{B}(L_p^N)$ we have 
$$
|f(i)|\le N^{1/p}=2^{dk/p}\le 2^{m^*/2-1}.
$$
Hence, by (\ref{E4}),
\begin{equation}
\label{N9}
\g(\b(f(i)))\le f(i)\le \g(\b(f(i)))+2^{-m^*/2}.
\end{equation}
Define 
$$
\Gamma:\mathcal{B}(L_p^N)\to W_p^r(D) \quad\mbox{by}
\quad\Gamma(f)=\sum_{i=0}^{N-1}\g\circ\b\circ f(i)\,\psi_i.
$$
By (\ref{N7}) and (\ref{N9}), for $f\in \mathcal{B}(L_p^N)$,
\begin{eqnarray*}
\|\Gamma(f)\|_{W_p^r(D)}&\le& \sigma_2 2^{rk}\|\g\circ\b\circ f\|_{L_p^N}\\ 
&\le& \sigma_2 2^{rk}\left(\|f\|_{L_p^N}+\|f-\g\circ\b\circ f\|_{L_p^N}\right)\\
&\le&\sigma_2 2^{rk}\left(1+2^{-m^*/2}\right).
\end{eqnarray*}
Furthermore, by (\ref{G7}),
\begin{eqnarray}
\label{ZA1}
I_d\circ \Gamma(f)&=&\sum_{i=0}^{N-1}\g\circ\b\circ f(i)I_d\psi_i\nonumber\\
&=&
\sigma_1 N^{-1}\sum_{i=0}^{N-1}\g\circ\b\circ f(i)\nonumber\\
&=&\sigma_1 S_N(\g\circ\b\circ f).
\end{eqnarray}
Define 
$$
\eta:D\to \Z[0,N)\quad \mbox{by}\quad \eta(s)=\min\{i\,|\,s\in D_{ki}\},
$$
with the $D_{ki}$ as in the beginning of the proof, and 
$$
\rho:D\times\Z[0,2^{m^*})\to\R \quad \mbox{by}\quad \rho(s,z)=\g(z)\psi_{\eta(s)}(s).
$$
Then
\begin{eqnarray*}
\Gamma(f)(s)&=&\sum_{i=0}^{N-1}\g\circ\b\circ f(i)\,\psi_i(s)\\
&=&\g\circ\b\circ f(\eta(s))\,\psi_{\eta(s)}(s)\\
&=&\rho(s,\b\circ f\circ\eta(s)).
\end{eqnarray*}
So $\Gamma$ is of the form (\ref{F1}) (with $\kappa=1$) and maps 
$$
\mathcal{B}(L_p^N)\quad
\mbox{into}\quad  \sigma_2 2^{rk}\left(1+2^{-m^*/2}\right)\mathcal{B}(W_p^r(D)).
$$
By Corollary \ref{cor:1} and Lemma 6(iii) in Heinrich (2001a),
\begin{eqnarray*}
e_{2n}^q(I_d\circ\Gamma,\mathcal{B}(L_p^N))
&\le& e_n^q\left(I_d,\sigma_2 2^{rk}\left(1+2^{-m^*/2}\right)\mathcal{B}(W_p^r(D))\right)\\
&=&\sigma_2 2^{rk}\left(1+2^{-m^*/2}\right)e_n^q(I_d,\mathcal{B}(W_p^r(D))).
\end{eqnarray*}
Using (\ref{N9}) again, we infer
$$
\sup_{f\in\mathcal{B}(L_p^N)} |S_Nf-S_N(\g\circ\b\circ f)|\le 2^{-m^*/2},
$$
and hence, by Proposition \ref{pro:1} and Lemma 6(i) and (ii) of Heinrich (2001a),
using also (\ref{ZA1}) and (\ref{N6}),
\begin{eqnarray*} 
cn^{-1}&\le &e_{2n}^q(S_N,\mathcal{B}(L_p^N))\\
&\le&  e_{2n}^q(S_N\circ\g\circ\b,\mathcal{B}(L_p^N))+2^{-m^*/2}\\
&=& \sigma_1^{-1}e_{2n}^q(I_d\circ\Gamma,\mathcal{B}(L_p^N))+2^{-m^*/2}\\
&\le& \sigma_1^{-1}\sigma_2 2^{rk}\left(1+2^{-m^*/2}\right)
e_n^q(I_d,\mathcal{B}(W_p^r(D)))+2^{-m^*/2}\\
&\le& c n^{r/d} e_n^q(I_d,\mathcal{B}(W_p^r(D)))+2^{-m^*/2}.
\end{eqnarray*}
Since $m^*$ can be made arbitrarily large, the desired result follows.
\end{proof}

\section{Comments}
Let us discuss the cost of the presented algorithm in the bit model of computation.
The algorithm consists of quantum summations on the levels $l=k_0,\dots,k-1$.
On level $l$ we have $N_l=2^{dl}$ and $n_l=\Theta(2^{-\delta(l-k_0)}n)$, where 
$\delta>0$ does not depend on $l$ or $n$. Recall also that $2^{dk_0}=\Theta(n)$
and $k-k_0=\Theta(\log n)$. 
Referring to the respective discussion of summation in the bit model in
Heinrich (2001a), section 6, and Heinrich and Novak (2001b), section 5, we conclude
that on level $l$ we need $\mathcal{O}(\log N_l)$ qubits, 
$\mathcal{O}(n_l \log N_l)$ quantum gates, $\mathcal{O}(\log n_l\,\log\log n_l)$
measurements in the case $p>2$ and 
$$
\mathcal{O}(n_l^2N_l^{-1}/\max(\log(n_l/\sqrt{N_l}),1)
+ \log (N_l/n_l)\log \log (N_l/n_l))
$$ 
measurements for $p\le 2$. Summarizing, we see that altogether the algorithm needs 
$\mathcal{O}(\log n)$ qubits, $\mathcal{O}(n \log n)$ quantum gates, 
$\mathcal{O}((\log n)^2\log\log n)$ measurements for $p>2$ and $\mathcal{O}(n/\log n)$
measurements for $p\le 2$.
 Thus the quantum bit cost differs by at most a logarithmic
factor from the quantum query cost $\Theta(n)$.

In the following table we summarize the results of this paper and compare them with
the respective known quantities of the classical deterministic and randomized setting.
We refer to Heinrich and Novak (2001a) and the bliography therein  for more information
on the classical setting. The respective entries of the table 
give the minimal error, constants and logarithmic factors are suppressed.

\[%\hspace*{-.5cm}
\begin{array}{l|l|l|l}
& \ \mbox{deterministic}\  & \, \mbox{random}\, & \, \mbox{quantum}\,\\ \hline 
&&&\\
\mathcal{B}(W_{p,d}^r(D)),\,2\le p< \infty &\, n^{-r/d } & \, n^{-r/d -1/2 }  
& \, n^{-r/d -1}\\
&&&\\
\mathcal{B}(W_{p,d}^r(D)),\,1< p<2 &\, n^{-r/d } & \, n^{-r/d -1 + 1/p }
&  n^{-r/d -1}\\ 
&&&\\
\mathcal{B}(W_{1,d}^r(D)) &\, n^{-r/d } & \, n^{-r/d }
&  n^{-r/d -1}\\ 
\end{array}
\]
The quantum rate for $1\le p<2$ is a certain surprise. Previous results 
led one to conjecture that the quantum setting could reduce the exponent
of the classical randomized setting by at most 1/2. 
Now we see that in the case $p=1$ there is even a reduction by 1.

\end{document}